# Interaction- and phonon-induced topological phase transitions in double helical liquids†

Chen-Hsuan Hsu

Helical liquids, formed by time-reversal pairs of interacting electrons in topological edge channels, provide a platform for stabilizing topological superconductivity upon introducing local and nonlocal pairings through the proximity effect. Here, we investigate the effects of electron–electron interactions and phonons on the topological superconductivity in two parallel channels of such helical liquids. Interactions between electrons in different channels tend to reduce nonlocal pairing, suppressing the topological regime. Additionally, electron–phonon coupling breaks the self duality in the electronic subsystem and renormalizes the pairing strengths. Notably, while earlier perturbative calculations suggested that longitudinal phonons have no effect on helical liquids themselves to the leading order, our nonperturbative analysis shows that phonons can induce transitions between topological and trivial superconductivity, thereby weakening the stability of topological zero modes. Our findings highlight practical limitations in realizing topological zero modes in various systems hosting helical channels, including quantum spin Hall insulators, higher-order topological insulators, and their fractional counterparts recently observed in twisted bilayer systems.

### New concepts

Helical liquids, a quantum state of matter appearing at the edges of time-reversal-invariant topological materials, hold promise for advanced electronics, including low-dissipation devices. When two parallel channels of these helical liquids are in contact with a conventional superconductor, they can stabilize topological zero modes without magnetic fields, providing a foundation for topological quantum computing. However, the robustness of these zero modes face challenges from detrimental effects such as Coulomb interactions and phonons. In this research, we show that both Coulomb interactions between electrons and electron–phonon coupling can lead to new topological phase transitions, weakening the stability of these zero modes. Using a combination of bosonization techniques and renormalization-group analysis, our investigation in the nonperturbative regime reveals new insights into these transitions, reaching a different conclusion from earlier perturbative studies. Since electron–electron interactions and phonons are omnipresent in nature, our findings not only open the door to electrically tunable topological transitions but also highlight significant practical constraints in designing stable devices for topological quantum computing.

Helical liquids, appearing at the boundary channels of time-reversal-invariant topological materials,[1–22] are promising candidates for advanced electronics such as low-dissipation devices, although the practical application presents certain complexities.[23–25] Along with their peculiar transport properties, these helical channels can facilitate Majorana and parafermion zero modes when proximity-induced pairing is introduced,[25–34] further extending previous proposals[35–39] exploiting nonhelical channels.[40–59] Notably, renewed interest in helical liquids has emerged following recent observations of double and triple quantum spin Hall states in twisted bilayer $MoTe_2$,[60] and $WSe_2$,[61] as well as dual quantum spin Hall states in monolayer $TaIrTe_4$,[62] including findings within the fractional regime.[60] There is also theoretical literature exploring fractional helical liquids in interacting systems.[29,63–68]

While the stability of the helical liquids themselves has been extensively analyzed,[5,6,25,69–101] it remains crucial to examine the robustness of the topological zero modes hosted therein against various detrimental effects in realistic settings. Among others, a crucial element in low dimensions is electron–electron interactions. Studies have demonstrated that Majorana zero modes can survive with the interactions[102] and, in certain settings, are even stabilized by them.[29,31,32,59] Additionally, the resilience of topological phases to phonons can pose challenges for experimental realization. Concerning helical liquids themselves, perturbative analyses have shown that the leading-order effect from longitudinal phonons on charge transport is identically zero.[72] Meanwhile, although transverse phonons could affect transport, any resulting corrections are suppressed by a high-power temperature law and vanish at low temperatures.[92] The influence of phonons on Majorana zero modes in

*Institute of Physics, Academia Sinica, Taipei 11529, Taiwan.*
*E-mail: chenhsuan@gate.sinica.edu.tw*

† Electronic supplementary information (ESI) available: The details about the model and the setup, operator product expansion, phonon influence, and additional numerical results of the RG analysis. See DOI: https://doi.org/10.1039/d4nh00254g

nonhelical systems has also been studied within the perturbative regime.[103]

In this work, we explore topological superconductivity in double helical liquids in regimes where both electron–electron interactions and electron–phonon coupling are nonperturbative. When the helical channels are in contact with a superconductor, both local and nonlocal pairing processes take place where the partners of a Cooper pair from the parent superconductor tunnel into one and both of the channels, respectively. Previous works[29,32] established that a topological regime is characterized by the dominance of nonlocal pairing over local pairing, a scenario that is feasible with sufficiently strong intrachannel electron–electron interactions. Here, we show that new transitions between topological and trivial phases can be induced by interchannel interactions. Specifically, while intrachannel interaction favors nonlocal pairing, interchannel interaction can suppress it. We further discuss the self duality in the electronic subsystem and the ground state degeneracy protected by parity conservation. Remarkably, we demonstrate that the coupling to phonons can break the self duality and alter the scaling behaviors of both pairing terms, thus triggering further phase transitions. Given the ubiquitous presence of electron–electron interactions and phonons, our findings suggest a practical limitation in realizing topological zero modes in helical channels.

Our setting with two parallel helical channels in proximity to an s-wave superconductor is illustrated in Fig. 1. In addition to configurations using quantum spin Hall insulators[29] or higher-order topological insulators[32] in the integer regime, we also consider edge channels in twisted bilayer MoTe$_2$,[60] which can be operated in either an integer or fractional regime upon adjusting the carrier density. The system can be described by $H = H_{dh} + H_{ph} + H_{ep} + V_{loc} + V_{cap}$. The first term, $H_{dh}$, describes electrons subject to intrachannel and interchannel screened Coulomb interactions. The next two, $H_{ph}$ and $H_{ep}$, describe the acoustic phonon modes and their coupling to the electrons, respectively. In this work, we treat the three terms nonperturbatively. Finally, we have perturbation terms, $V_{loc}$ and $V_{cap}$, arising from the local and nonlocal pairings, respectively. Below, we first discuss the electronic subsystems described by $H_{dh} + V_{loc} + V_{cap}$, before analyzing the influence of phonons.

In our analysis, the electrons are represented by the operator $\psi_{\ell,n}(r)$ with the spatial coordinate $r$, the channel index $n \in \{1, 2\}$, and $\ell = R$ ($\ell = L$) for right-moving spin-down (left-moving spin-up) modes, respectively. Within the bosonization scheme,[63–65] we have

$$\psi_{\ell,n}(r) = \frac{U_{\ell,n}}{\sqrt{2\pi a}} e^{i\ell k_F r} e^{im[\ell\phi_n(r)+\theta_n(r)]}, \quad (1)$$

with the Fermi wave number $k_F$, short-distance cutoff $a = \hbar v_F/\Delta_a$, bandwidth $\Delta_a$ and Fermi velocity $v_F$. Here, the boson fields satisfy

$$[\phi_n(r), \theta_{n'}(r')] = \frac{i\pi}{2m}\delta_{nn'}\text{sign}(r' - r), \quad (2)$$

with integer $m$. We have the following bosonized expression characterizing double helical liquids with nonuniversal forward scattering interactions,

$$H_{dh} = \sum_\delta \int dr \frac{\hbar u_\delta}{2\pi}\left[\frac{1}{K_\delta}(\partial_r\phi_\delta)^2 + K_\delta(\partial_r\theta_\delta)^2\right], \quad (3)$$

with the index $\delta \in \{s \equiv +, a \equiv -\}$ labeling the symmetric/antisymmetric combination of the two channels, $\phi_\delta \equiv (\phi_1 + \delta\phi_2)/\sqrt{2}$ and similarly for $\theta_\delta$. The velocity $u_\delta = v_F/K_\delta$ of the sector $\delta$ is related to the interaction parameter $K_\delta = [1 + 2(U_{ee} + \delta V_{ee})/(\pi\hbar v_F)]^{-1/2}$ with the intrachannel (interchannel) interaction strength $U_{ee}$ ($V_{ee}$). One obtains compactness in the boson fields, $\phi_\delta \sim \phi_\delta + \frac{2\sqrt{2}\pi}{m}$ and $\theta_\delta \sim \theta_\delta + \frac{2\sqrt{2}\pi}{m}$, which plays a role when examining the ground state degeneracy below.

The proximity-induced local and nonlocal pairings (characterized by the strengths $\Delta_n$ with $n \in \{1,2\}$ and $\Delta_c$, respectively) can be expressed as[25,29,32]

$$V_{loc} = \int dr \frac{2\Delta_+}{\pi a}\cos(\sqrt{2}m\theta_s)\cos(\sqrt{2}m\theta_a)$$
$$- \int dr \frac{2\Delta_-}{\pi a}\sin(\sqrt{2}m\theta_s)\sin(\sqrt{2}m\theta_a), \quad (4)$$

$$V_{cap} = \int dr \frac{2\Delta_c}{\pi a}\cos(\sqrt{2}m\theta_s)\cos(\sqrt{2}m\phi_a), \quad (5)$$

with $\Delta_\pm = (\Delta_1 \pm \Delta_2)/2$. In the noninteracting limit, the band inversion and topological phase transition occurs when

$$\Delta_c^2 + \Delta_-^2 > \Delta_+^2, \quad (6)$$

which gives the criterion for the emergence of Majorana zero modes.[29,32] For simplicity, below we consider the identical pairing setup with $\Delta_- = 0$.‡ As demonstrated below, in this case, one recovers a self-dual sine-Gordon model;[105] for $m = 2$, the model coincides with two-leg ladder systems,[106,107] which host topologically nontrivial pair density wave phases, albeit with distinct realizations. The criterion given in eqn (6) was derived from the single-particle Hamiltonian.[29,32] To gain more insights into the many-body systems, below we revisit it with the bosonic description.

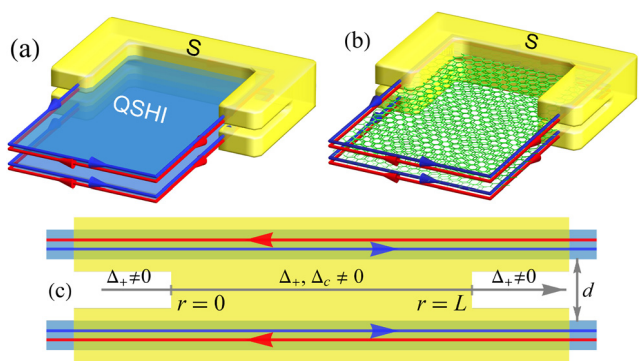

Fig. 1 Double helical liquids realized using (a) quantum spin Hall insulators (QSHI) or (b) twisted bilayer structures. In both settings, a pair of parallel edge channels, each formed by counter-propagating spin-up (red) and spin-down (blue) electrons, are in contact with an s-wave superconductor (yellow). (c) Side view of the setup in the local coordinate (labeled by r), with an interlayer separation d. The structure induces both local and nonlocal pairings within the segment $0 \le r \le L$, whereas only local pairing occurs for $r < 0$ and $r > L$.

We now explore the ground state degeneracy determined by the parity conservation, a method shown to be effective in unveiling the topological properties of semiconducting wires[108] and spin chains.[107,109,110] We note that, owing to the presence of nonlocal pairing, the fermion parity for a single channel here is not conserved. However, since the total fermion parity defined from the total charge remains conserved, we formulate it as $P_f = (-1)^{q_1+q_2} = e^{-\sqrt{2}i \int dr \partial_r \phi_s}$ with the fermion number $q_n$ for channel $n$. Utilizing the relation $P_f \theta_s P_f^{-1} = \theta_s - \sqrt{2}\pi/m$ and the compactness of $\theta_s$, we conclude that the ground states for $\cos(\sqrt{2}m\theta_s)$ are characterized by eigenstates of $P_f$, satisfying $P_f|e/o\rangle_s = \pm|e/o\rangle_s$. These states are constructed from even and odd combinations as $|e/o\rangle_s = \left(|\theta_s = \theta_0\rangle_s \pm |\theta_s = \theta_0 - \sqrt{2}\pi/m\rangle_s\right)/\sqrt{2}$, with $\theta_0$ being a minimum of the cosine term. Concerning the topological properties, since the $\theta_s$ field appears in both $V_{\text{loc}}$ and $V_{\text{cap}}$, the total fermion parity does not distinguish between the two terms.

In contrast, since $V_{\text{loc}}$ and $V_{\text{cap}}$ involve different fields, $\phi_a$ and $\theta_a$, in the antisymmetric sector, one can exploit another parity operator to distinguish between the two. Specifically, given the singlet pairing considered here, both local and nonlocal pairings create or annihilate up- and down-spin pairs simultaneously. As a result, the pairings can alter the spin difference between the two channels by either 0 or 2 (in units of $\hbar/2$), thereby preserving a quantity, which we term the "spin difference parity". This leads to the definition of an operator $P_{\text{sp}}$ using the spin quantum number $s_n$ for channel $n$, $P_{\text{sp}} = (-1)^{s_1-s_2} = e^{-\sqrt{2}i \int dr \partial_r \theta_a}$. Again, with $P_{\text{sp}} \phi_a P_{\text{sp}}^{-1} = \phi_a - \sqrt{2}\pi/m$ and the compactness of $\phi_a$, we conclude that the ground states for $\cos(\sqrt{2}m\phi_a)$ are doubly degenerate and can be written as $|e/o\rangle_a = \left(|\phi_a = \phi_0\rangle_a \pm |\phi_a = \phi_0 - \sqrt{2}\pi/m\rangle_a\right)/\sqrt{2}$, with eigenvalue $\pm 1$ of $P_{\text{sp}}$ and a minimum $\phi_0$. On the other hand, since the spin difference parity operator commutes with $\theta_a$, the local pairing term exhibits no ground state degeneracy. We thus conclude that the $\Delta_c$ term characterizes a topologically nontrivial phase.§

Having identified that a dominant $\Delta_c$ term leads to topological superconductivity, we explore its resilience to both intra-channel and interchannel electron–electron interactions. Under these conditions, the criterion in eqn (6) holds upon replacing the pairing strengths to their renormalized values by the interactions.[32,102] To proceed, we perform the renormalization-group (RG) analysis and obtain these renormalized strengths; see ESI† for the RG flow equations without phonon contribution (ESI†). In Fig. 2, we present the results for $m = 1$. The influence of electron–electron interactions is notable. A general trend toward dominant nonlocal pairing requires sufficiently large $U_{ee}$ to preferentially suppress local pairing over nonlocal ones, consistent with earlier findings.[32] In contrast, finite interchannel interaction leads to a larger $K_a$ value and reduces the nonlocal pairing. A sufficiently large $V_{ee}$ eventually induces a phase transition toward the trivial superconductivity, suppressing formation of topological zero modes.

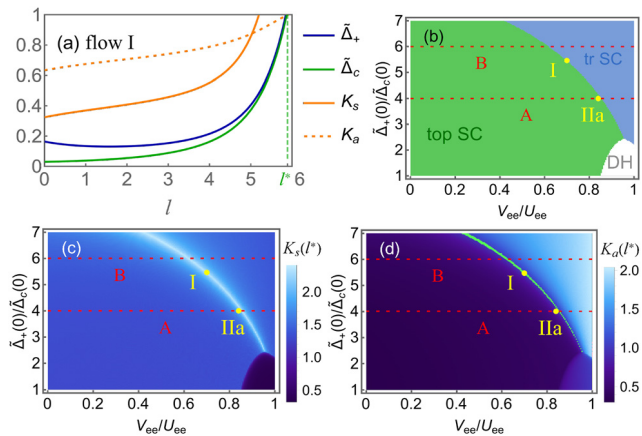

Fig. 2 RG flow and phase diagrams in the absence of phonons. (a) RG flow for $\tilde{\Delta}_+(0) = 0.164$, $\tilde{\Delta}_c(0) = 0.03$, $U_{ee}/(\pi\hbar v_F) = 2.5$ and $V_{ee}/U_{ee} = 0.7$. The label $l^*$ marks the length scale of the flow end point, where any of the couplings reach unity. (b) Phase diagram determined from the renormalized pairing strengths, with the labels "top SC", "tr SC" and "DH" denoting topological superconductivity, trivial superconductivity and double helical liquids, respectively. In the last region, none of the pairings reaches unity as the scale $l$ approaches $\ln(L/a)$ set by the system size $L$. (c) and (d) Renormalized interaction parameters, with a highlighted curve where $K_a$ flows to unity. The initial points of the flows in panel (a) and Fig. 3(a) are marked; the line cuts A and B correspond to those in Fig. 4.

Interestingly, both local and nonlocal pairings tend to increase $K_s$, thereby promoting the ordering of the $\theta_s$ field. Assuming that the $\theta_s$ field is frozen, the low-energy physics is primarily governed by the antisymmetric sector. We can express the pairing terms $V = V_{\text{loc}} + V_{\text{cap}}$ as

$$V = \int \frac{dr}{a}\left[g_\phi \cos(\lambda_\phi \Phi) + g_\theta \cos(\lambda_\theta \Theta)\right], \quad (7)$$

with $g_\phi \propto \Delta_c \langle \cos(\sqrt{2}m\theta_s)\rangle$, $g_\theta \propto \Delta_+ \langle \cos(\sqrt{2}m\theta_s)\rangle$, $\lambda_\phi = \sqrt{2\pi m K_a}$ and $\lambda_\theta = \sqrt{2\pi m/K_a}$. The new fields $\Phi = \sqrt{m/(\pi K_a)}\phi_a$ and $\Theta = \sqrt{m K_a/\pi}\theta_a$ are introduced to match the notations in ref. 105. We observe that the electronic subsystem characterizes a self-dual sine-Gordon Hamiltonian, with the self-duality point $g_\phi = g_\theta$ and $\lambda_\phi = \lambda_\theta$ reached when $K_a \to 1$. Remarkably, for certain initial parameters, such as flow I in Fig. 2(a), the renormalized value of $K_a$ converges to unity at the end of the RG flow, accompanied by $\tilde{\Delta}_s(l^*) = \tilde{\Delta}_a(l^*)$, signifying the arrival at the self-duality point. Moreover, we find that eqn (7) presents a hierarchy of self-dual sine-Gordon models (ESI†), where different $m$ values characterize different classes with $\lambda_\phi^2 = \lambda_\theta^2 = 2\pi m$,[105,111,112] including the $m = 3$ model with parafermion modes. Notably, for $m = 2$, eqn (7) can be refermionized using $\Psi_\ell = e^{\sqrt{\pi}i(-\ell\Phi+\Theta)}$, allowing for the introduction of Majorana fields $\xi_\ell$ and $\eta_\ell$ through $\Psi_\ell = (\xi_\ell + i\eta_\ell)/\sqrt{2}$. With the details in the ESI,† we get

$$V = i \int \frac{dr}{2a} \left[ (g_\phi + g_\theta) \xi_R \eta_L + (g_\theta - g_\phi) \eta_R \xi_L \right], \quad (8)$$

indicating a pair of massive Majorana modes $\xi_R$ and $\eta_L$ for any $g_\phi$ and $g_\theta$. In contrast, the $\eta_R$ and $\xi_L$ modes can be massless at a junction where $g_\theta$ and $g_\phi$ reverse their relative strength. Translating back to the original notations, this indicates a transition point at which $(\Delta_c - \Delta_+)$ changes its sign and implies topological zero modes at $r = 0, L$ in Fig. 1 with a dominant nonlocal pairing for $0 \leq r \leq L$, recovering the criterion obtained earlier.

We have examined the robustness of the topological superconductivity in the proximitized double helical liquid, which exhibits self duality in the electronic subsystems and interaction-driven phase transition. However, phonons, generally present in nature, lead to a breakdown of this duality and an additional phase transition, as we discuss below.

To explore the phonon effects, we consider the following term from longitudinal acoustic phonons,

$$H_{ph} = \sum_n \int \frac{dr}{2\rho} \left[ \pi_n^2 + \rho^2 c^2 (\partial_r d_n)^2 \right], \quad (9)$$

with the phonon velocity $c$, the linear mass density $\rho$ along the channels, the displacement field $d_n$ generated by the phonons and its conjugate field $\pi_n$. The coupling between the displacement field and the electron density is described by

$$H_{ep} = \sum_n g \int dr (\partial_r \phi_n)(\partial_r d_n), \quad (10)$$

with the electron–phonon coupling strength $g$. We remark that isotope engineering can be employed to modify the phonon parameters, as the atomic mass of the lattice can be altered through isotopic substitution, thus affecting the vibrational properties of the lattice. In earlier perturbative analysis,[72] the phonons introduced above are shown to have no leading-order effects on the helical liquids themselves. However, nonperturbative calculation in nonhelical systems[113–116] suggests that phonons can affect the scaling dimensions of pairing operators, which in turn likely affects the phase diagram that we are examining here.

Before proceeding, we comment on optical phonons. With their weakly dispersive energy band near a finite frequency, they have a negligible phonon density of states at low energy compared to acoustic modes,[117] leading to reduced contributions at low temperatures. Additionally, these phonons are not expected to produce substantial effects on the scaling dimensions.[115] We therefore follow the literature on one-dimensional bosonized models[72,113–116] and do not include their contributions here.

From eqn (10), it is evident that phonons couple to the $\phi_n$ fields but not to the $\theta_n$ fields, leading to a breakdown of the self duality. Moreover, due to the hybridization of electrons and phonons, the low-energy modes of $H_{dh} + H_{ph} + H_{ep}$ are characterized by velocities,

$$u_{\delta,\eta} = \left( \frac{u_\delta^2 + c^2}{2} + \frac{\eta}{2} \sqrt{(u_\delta^2 - c^2)^2 + 4 v_g^4} \right)^{1/2}, \quad (11)$$

with $\eta \in \{+, -\}$ and $v_g \equiv \sqrt[4]{\frac{\pi v_F g^2}{\hbar \rho}}$. The modification in velocities from their bare values, $u_\delta$ and $c$, can be substantial, to the extent that one of the velocities reaches zero, which is referred to as the Wentzel–Bardeen (WB) singularity, known for non-helical channels.[113–116,118–120] Without pairing terms, the singularity condition is quantified as

$$\frac{v_g^4}{c^2 v_F^2} = 1 + \frac{2}{\pi \hbar v_F}(U_{ee} - V_{ee}), \quad (12)$$

implying that the inclusion of interchannel interaction makes it easier for the system to reach the singularity. The presence of eqn (4) and (5) results in the renormalization of the interaction parameters and a more complicated condition than eqn (12); we therefore determine the singularity condition numerically. Crucially, the electron–phonon coupling also modifies the scaling dimensions of the pairing operators, thereby affecting the phase diagrams.

To quantify the effects from phonons, we derive the following RG flow equations,

$$\frac{d\tilde{\Delta}_+}{dl} = \left[ 2 - \frac{m}{2} \sum_{\eta = \pm} \left( \frac{u_s \gamma_{s,\eta}^\theta}{K_s u_{s,\eta}} + \frac{u_a \gamma_{a,\eta}^\theta}{K_a u_{a,\eta}} \right) \right] \tilde{\Delta}_+, \quad (13a)$$

$$\frac{d\tilde{\Delta}_c}{dl} = \left[ 2 - \frac{m}{2} \sum_{\eta = \pm} \left( \frac{u_s \gamma_{s,\eta}^\theta}{K_s u_{s,\eta}} + \frac{u_a K_a \gamma_{a,\eta}^\phi}{u_{a,\eta}} \right) \right] \tilde{\Delta}_c, \quad (13b)$$

$$\frac{dK_s}{dl} = 2m\tilde{\Delta}_+^2 + 2m\tilde{\Delta}_c^2, \quad (13c)$$

$$\frac{dK_a}{dl} = 2m\tilde{\Delta}_+^2 - 2mK_a^2 \tilde{\Delta}_c^2, \quad (13d)$$

with $dl = da/a$, $\tilde{\Delta}_+ = \Delta_+/\Delta_a$, $\tilde{\Delta}_c = \Delta_c/\Delta_a$ and

$$\gamma_{\delta,\eta}^\phi = \eta \left( \frac{u_{\delta,\eta}^2 - c^2}{u_{\delta,+}^2 - u_{\delta,-}^2} \right), \quad (14a)$$

$$\gamma_{\delta,\eta}^\theta = \frac{\eta}{u_\delta^2} \left( \frac{u_\delta^2 u_{\delta,\eta}^2 - u_\delta^2 c^2 + v_g^4}{u_{\delta,+}^2 - u_{\delta,-}^2} \right). \quad (14b)$$

The equations indicate that the electron–phonon coupling affects the scaling of operators associated with both local and nonlocal pairings. In Fig. 3, we compare the RG flows for scenarios with and without phonon influence. In both instances, $K_s$ is renormalized in a similar manner, supporting both pairing types. The key distinction arises with phonons, where $K_a$ is renormalized towards larger values, favoring local pairing over the nonlocal one. Notably, the RG flows with and without phonons result in the opposite outcomes, despite otherwise identical initial parameter values.

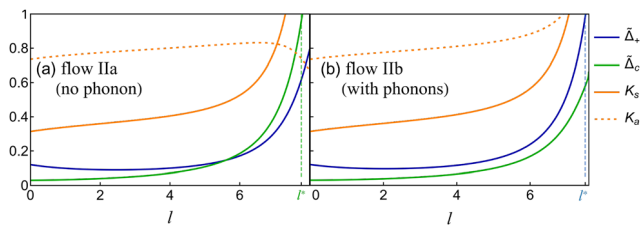

Fig. 3 Comparison of RG flows without and with phonon influence for $\tilde{\Delta}_+(0) = 4\tilde{\Delta}_c(0) = 0.12$, $U_{ee}/(\pi\hbar v_F) = 2.5$ and $V_{ee}/U_{ee} = 0.83$. (a) RG flow corresponding to dot IIa in Fig. 2 and 4(a). (b) RG flow with $v_g^2/(cv_F) = 0.53$, corresponding to dot IIb in Fig. 4(a).

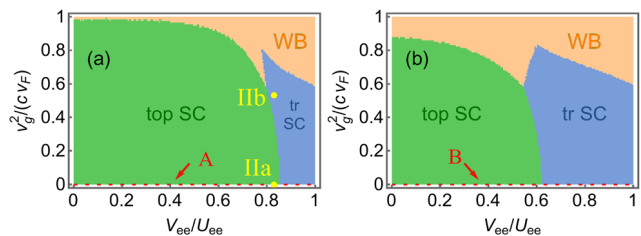

Fig. 4 Phase diagrams in the presence of phonons with $U_{ee}/(\pi\hbar v_F) = 2.5$ for (a) $\tilde{\Delta}_+(0) = 4\tilde{\Delta}_c(0) = 0.12$ and (b) $\tilde{\Delta}_+(0) = 6\tilde{\Delta}_c(0) = 0.18$. The label "WB" indicates the regime beyond the WB singularity. The dots IIa and IIb correspond to the flows in Fig. 3(a) and (b), respectively. The line cuts A and B correspond to those marked in Fig. 2.

Next, we deduce phase diagrams by exploring a range of the dimensionless parameters, $v_g^2/(cv_F)$ and $V_{ee}/U_{ee}$. As illustrated in Fig. 4, the presence of electron–phonon coupling significantly alters the phase diagrams. Specifically, since phonons mediate attractive interactions between electrons within each channel, their presence effectively reduces the strength of $U_{ee}$. In terms of the RG flow, a nonzero $v_g$ increases both the $K_s$ and $K_a$ values, thereby enhancing $\tilde{\Delta}_+$ and reducing $\tilde{\Delta}_c$. In consequence, close to the boundary between the two superconducting phases, the electron–phonon coupling can push the system from a topological phase to a trivial phase. Furthermore, the pairing perturbations also influence the boundary approaching the WB singularity. Consequently, the WB singularity is reached for a smaller $v_g$ as compared to earlier analysis.[113,114] We also note that, for $m > 1$, a weaker electron–phonon coupling is necessary to induce phase transitions (ESI†), indicating a more fragile topological phase in the fractional regime.

Our results indicate electrically tunable topological phase transitions in proximitized double helical liquids upon varying the strengths of $U_{ee}$ and $V_{ee}$. As demonstrated in the ESI,† the intrachannel strength $U_{ee}$ can be modulated through the screening effect,[32,69,121] which is controlled by either the distance between the channel and a nearby metallic layer or the dielectric material in between. As the interchannel strength $V_{ee}$ additionally depends on the distance $d$ and the dielectric material between the two helical channels,[116] they offer extra knobs to adjust the ratio of $V_{ee}/U_{ee}$ through device engineering. In consequence, the electrical tunability of the proposed setups allows one to induce the phase transitions, which are associated with the presence or absence of topological zero modes. The latter feature can be detected through various probes,[34] including transport measurements,[48,52,122] magnetic field responses,[49,51,52] and the parity-controlled $2\pi$-Josephson effect previously proposed.[123,124] Given the omnipresence of electron–electron interactions and phonons, even in clean systems, our findings represent practical constraints in utilizing helical channels to realize topological zero modes.

## Data availability

All data that support the findings of this work are included within the article and ESI.†

## Conflicts of interest

There are no conflicts to declare.

## Acknowledgements

We thank S.-Y. Chen, M. Hayashi, Y. Kato, T. Kondo, and R. Noguchi for interesting discussions. This work was supported financially by the National Science and Technology Council (NSTC), Taiwan through Grant No. NSTC-112-2112-M-001-025-MY3. We are also grateful to the long-term workshop YITP-T-23-01 held at the Yukawa Institute for Theoretical Physics (YITP), Kyoto University, Japan, where this work was initiated.

## Notes and references

‡ To achieve the criterion (6), one can alternatively consider the $\pi$-junction setup with $\Delta_+ = 0$ but finite $\Delta_-$,[30,104] which can be mapped back to the same form as the identical pairing setup (ESI†).

§ Despite a similar form, $\cos(\sqrt{4\pi}\phi)$ and $\cos(\sqrt{4\pi}\theta)$, in ref. 107, the roles of the two cosine terms are reversed: the topological phase is characterized by their current-like $\theta$ field, since the parity is related to their density-like $\phi$ field.